\documentclass[11pt,a4paper]{article}
\usepackage{jheppub}
\usepackage{amsthm}
\usepackage{mathtools}
\usepackage{physics}
\usepackage{array}
\usepackage{enumitem}
\usepackage{bm}
\usepackage{hyperref}
\usepackage[capitalize]{cleveref}
\usepackage{graphicx}
\usepackage[vcentermath]{youngtab}
\newcommand{\Q}{\mathbb{Q}}
\newcommand{\C}{\mathbb{C}}
\newcommand{\R}{\mathbb{R}}
\newcommand{\Z}{\mathbb{Z}}
\arxivnumber{2501.09860}

\title{More varieties of 4-d gauge theories: product representations}
\author[a]{Ben Gripaios}
\author[b]{and Khoi {Le Nguyen Nguyen}}

\affiliation[a]{Cavendish Laboratory, University of Cambridge, J.J. Thomson Avenue, Cambridge, CB3 0HE, United Kingdom}

\affiliation[b]{DAMTP, University of Cambridge, Wilberforce Road, Cambridge, CB3 0WA, United Kingdom}

\emailAdd{gripaios@hep.phy.cam.ac.uk}
\emailAdd{kl518@cam.ac.uk}
\newcolumntype{L}{>{$}l<{$}} 
\newcolumntype{R}{>{$}r<{$}} 

\abstract{Recently, we used methods of arithmetic geometry to study the anomaly-free irreducible representations of an arbitrary gauge Lie algebra. Here we generalize to the case of products of irreducible representations, where it is again possible to give a complete description. A key result is that the projective variety corresponding to $m$-fold product representations of the Lie algebra $\mathfrak{su}_n$ is a rational variety for every $m$ and $n$. We study the simplest case of $\mathfrak{su}_3$  (corresponding to the strong interaction) in detail. We also describe the implications of a number-theoretic conjecture of Manin (and related theorems) for the number of chiral representations of bounded size $B$ (measured roughly by the Dynkin labels) compared to non-chiral ones, giving a precise meaning to the sense in which the former (which are those most relevant for phenomenology) are rare compared to the latter. As examples, we show that, for both irreducible representations of $\mathfrak{su}_5$ and once-reducible product representations of $\mathfrak{su}_3$ that are non-anomalous, the number of chiral representations is asymptotically between $B (\log B)^5$ and $B^\frac{4}{3}$, while the number of non-chiral representations is asymptotically $B^2$. Despite this rarity of chiral, anomaly-free, product representations, we show that there are examples relevant for phenomenology, including one that gives an asymptotically-free gauge theory with Lie algebra $\mathfrak{su}_7$.} 

\begin{document}
	\maketitle
	\flushbottom
	\section{Introduction \label{sec:intro}}
	For a gauge theory to be consistent up to high energies, anomalies must cancel, putting non-trivial constraints on the allowed fermionic matter content. 
	The idea of applying concepts of arithmetic geometry to the problem of cancellation of anomalies in four-dimensional gauge theories that are local ({\em i.e.} those which depend only on the datum of the Lie algebra of the gauge theory rather than its Lie group) was introduced in \cite{Allanach_2020}, where it was used to 
	study arbitrary representations of the simplest 
	Lie algebra, namely $\mathfrak{u}_1$. Recently, it was used \cite{Gripaios_2024} to study the simplest representations, namely the irreducible ones, of an arbitrary Lie algebra.\footnote{For an application to the Standard Model, see \cite{Allanach_2020_3}, and for an application to 2-d gauge theories, see \cite{Camp_2024}.}
	
	These two cases are complementary, in that the only example which is common to both is the trivial representation of $\mathfrak{u}_1$ (since any other irreducible representation of an abelian Lie algebra is necessarily anomalous). Given that, we are emboldened to go further and ask whether it can be applied to other cases, with reducible representations of some Lie algebra other than $\mathfrak{u}_1$.
	
	Here we study a case that is a generalization of \cite{Gripaios_2024}, namely where we have an arbitrary Lie algebra $\mathfrak{g}$, but we allow the fermions to transform in a representation that is an $m$-fold (tensor) product of irreducible representations (henceforth: irreps), for some $m$. This case turns out to be a relatively straightforward generalization of the $m=1$ case, namely irreps of $\mathfrak{g}$.
	
	The generalization to products of irreps, whilst mathematically relatively straightforward, has significant implications for physics. Indeed, a drawback of the anomaly-free chiral irreps that we found in \cite{Gripaios_2024} is that they are typically very large as measured by, {\em e.g.}, their dimension or their largest Dynkin label. To give one example, the smallest anomaly-free chiral irrep of $\mathfrak{su}_5$ has dimension $1\, 357\, 824$, but there are 35 non-chiral irreps of smaller dimension. To give another, the smallest anomaly-free chiral once-reducible rep of $\mathfrak{su}_3$ has dimension $4\, 312$, but there are 41 non-chiral once-reducible representations of smaller dimension. As such, one might expect a QFT including fermionic matter in such a representation to rapidly become strongly-coupled as we move to higher energies, leading to the theory being predictive in only a narrow window of energy scales.\footnote{We thank the referee for this observation.} By generalizing to products of irreps, we obtain chiral anomaly-free representations that are much smaller. In fact, there is even an example, with Lie algebra $\mathfrak{su}_7$, giving an asymptotically-free QFT, giving a theory of physics that is valid up to arbitrarily high energies.
	
	Getting back to the mathematical story, we show that, just as for the $m=1$ case, we obtain a projective variety over the field of rational numbers $\mathbb{Q}$ that is, moreover, a rational variety (meaning that it is birationally equivalent to projective space). This makes it possible to not only merely `solve' the equations (which means, since there usually infinitely many solutions, to give a parameterization of them, and had already been achieved for the $\mathfrak{u}_1$ case in \cite{Costa_2019}), but to give a detailed geometric picture of them as well.
	
	To get a more concrete idea of what we mean by this, the reader might glance ahead to Fig.~\ref{fig:v32productrationalpoints}, which shows the rational points corresponding to anomaly-free representations for an example we discuss in exhaustive detail, namely the case of binary products (so $m=2$) of irreps of the Lie algebra $\mathfrak{su}_3$.\footnote{This variety has been studied before in mathematics under the name of the \emph{equianharmonic cubic surface} \cite{Segre_1942}.} This case has some relevance for phenomenology, since $\mathfrak{su}_3$ describes the strong interactions. In Nature, we currently observe only a non-chiral representation of $\mathfrak{su}_3$, given by the six flavours of quarks and their antiquarks, and there are no chiral anomaly-free irreps. But as we will see, there are plenty of binary (and higher) products of irreps that are both chiral and anomaly-free, and which could therefore play a r\^{o}le in physics beyond the Standard Model.
	
	The geometric approach also allows us to gain some further insights that are relevant for phenomenology. As we already observed for the case of irreps in \cite{Gripaios_2024}, chiral anomaly-free representations (which are needed to build natural theories with light fermions) are infinite in number (in cases where they exist at all) and even dominate the non-chiral representations in a specific geometric sense (namely, the chiral ones are dense, in either the usual euclidean topology or the Zariski topology, in the underlying real variety, while the non-chiral ones are not). Yet, as we have discussed, chiral anomaly-free representations tend to have very large dimensions compared to non-chiral ones.
	
	This motivates studying the question of counting chiral {\em vs.} non-chiral solutions of bounded size. As it turns out, this too can be turned into a question that can be attacked (if not yet completely answered) geometrically. As we shall see, an easy argument shows that the number of non-chiral solutions grows asymptotically as $B^2$ (where $B$ is some measure of the size of the Dynkin labels to be defined later), while a conjecture of Manin \cite{Franke_1989} here boils down to the claim that the number of chiral solutions grows as $B (\log B)^{\rho -1}$, where $\rho$ is the rank of the Picard group of the corresponding variety. Whilst Manin's conjecture has yet to be verified for any of the varieties we consider, there are a number of relevant bounds. For example, for both irreps of $\mathfrak{su}_5$ and binary products of irreps of $\mathfrak{su}_3$, we can carry over existing results from number theory \cite{HeathBrown_1997} to show that the number of chiral solutions is bounded above by $B^{4/3}$. (In both cases, Manin's conjectured number has also been shown \cite{Slater_1998} to be a lower bound.) So once we put a bound on the size of representations, as presumably we should when building physics models, we see that chiral anomaly-free product representations, even when they are dense, are few-and-far-between. Nevertheless, the asymptotically-free example based on $\mathfrak{su}_7$ mentioned above shows that they can be relevant for phenomenology, and we give some other examples.
	
	The outline is as follows. In Section~\ref{sec:gen} we show that for products of irreps of an arbitrary $\mathfrak{g}$, the constraints imposed by anomaly cancellation define a projective variety (more specifically a cubic hypersurface) and that this variety is rational. In Section~\ref{sec:bin3} we discuss the specific case of binary products of irreps of $\mathfrak{su}_3$ in detail. In Section~\ref{sec:moreprod} we give a selection of results for other cases. In Section~\ref{sec:pheno} we begin our discussion of the relevance for phenomenology. In the final Section we discuss the distribution of solutions via Manin's conjecture and related theorems. 
	
	\section{Generalities of product representations \label{sec:gen}}
	The case of anomaly-free irreps of an arbitrary gauge Lie algebra $\mathfrak{g}$ was studied in \cite{Gripaios_2024}; we now wish to generalize to representations that are products of irreps. Given a compact Lie algebra $\mathfrak{g}$ and a representation $\rho$ thereof, the anomaly cancellation equations read
	\begin{gather} \label{eq:ano}
		\mathrm{tr} \rho = \mathrm{tr} \rho^3 = 0.
	\end{gather}
	
	As in \cite{Gripaios_2024}, in solving these we are helped by a number of serendipitous facts.
	
	The first of these is that every product of irreps of an abelian algebra is one-dimensional and that every such representation is anomalous unless it is the trivial representation. As a result, we need consider only the semisimple summand of $\mathfrak{g}$.
	
	The second of these is that for such algebras $ \mathrm{tr} \rho = 0$ holds automatically and that $\mathrm{tr} \rho^3$ decomposes into a sum of contributions from each of its simple summands.
	
	The third is that $\mathrm{tr} \rho^3=0$ holds automatically for all simple Lie algebras except $\mathfrak{su}_n$, with $n \geq 3$. 
	
	The fourth is that in such cases, we can express $\mathrm{tr} \rho^3$ for an irrep in a particularly simple form. Namely, we can replace the usual Dynkin labels $(m_1,\dots,m_{n-1})$ (whose entries are nonnegative integers) with either the $n-1$ positive integers $(q_1,\dots,q_{n-1}):=(m_1+1,\dots,m_{n-1}+1)$ or the 
	$n$ integers 
	\begin{equation}
		\sigma_i:=-\sum_{k=1}^{i-1}kq_k+\sum_{k=i}^{n-1}(n-k)q_k,\label{eqn:sigma_from_q}
	\end{equation}
	satisfying the constraint
	\begin{equation} \label{eq:sigmalin}
		\sum_{i=1}^{n}\sigma_i=0
	\end{equation}
	along with
	\begin{equation}
		\sigma_i-\sigma_{i+1}=nq_i\in n\mathbb{Z}^+,i\in\{1,\dots,n-1\}. \label{eqn:sigma_order}
	\end{equation}
	Doing so, the anomaly takes the form \cite{Georgi_1976, Okubo_1977},
	\begin{equation}
		A(\rho)=\frac{2}{n^2(n^2-1)(n^2-4)}D(\rho)\sum_{i=1}^{n}(\sigma_i)^3,
	\end{equation}
	where the dimension is given by
	\begin{equation} D(\rho)=\prod_{j=1}^{n-1}\left[\frac{1}{j!}\prod_{k=j}^{n-1}\left(\sum_{i=k-j+1}^kq_i\right)\right]=\frac{\prod_{j<k}^n(\sigma_j-\sigma_k)}{n^{n(n-1)/2}\prod_{i=1}^{n-1}i!}.
	\end{equation}
	These are homogeneous polynomials (in either set of variables $\{q_i\}$ or $\{\sigma_j\}$, but not in the Dynkin labels).
	
	The fifth is that the usual formul\ae\ for the anomaly and dimension of a product of representations, {\em i.e.}
	\begin{align}
		D\left(\rho_1\otimes\rho_2\right)&= D\left(\rho_1\right) D\left(\rho_2\right), \\
		A\left(\rho_1\otimes\rho_2 \right)&= A\left(\rho_1\right) D\left(\rho_2 \right) + D\left(\rho_1\right)A\left(\rho_2\right),
		\label{eqn:product_two_reps}
	\end{align}
	which generalize to an $m$-fold product in an obvious way, show that each term in the formula for a product of $m$ irreps contains a common factor of the product of the dimensions of each of the irreps. Since this factor cannot vanish (the dimension of an irrep is positive-definite), our problem reduces to solving a single equation of rather low degree, namely the cubic
	\begin{equation}
		\sum_{\alpha=1}^{m}\sum_{i=1}^{n}(\sigma_{i\alpha})^3=0, \label{eqn:sigma3}
	\end{equation}
	where the additional subscript $\alpha$ indicates that the $n$ values $\sigma_{1\alpha},\dots,\sigma_{n\alpha}$ are those characterising the $\alpha$-th irrep factor ($\alpha \in \{1,\dots,m\}$) of the product, along with the
	$m$ linear constraints
	\begin{equation}
		\sum_{i=1}^n\sigma_{i\alpha}=0. \label{eqn:sigma1}
	\end{equation}
	Here, the $q$-tuples of each irrep, $(q_{1\alpha},\dots,q_{n-1,\alpha}):=(m_{1\alpha}+1,\dots,m_{n-1,\alpha}+1)$ satisfy the following relations for $i\in\{1,\dots,n-1\}$ and $\alpha\in\{1,\dots,m\}$:
	\begin{align}
		\sigma_{i\alpha}=-\sum_{k=1}^{i-1}kq_{k\alpha}+\sum_{k=i}^{n-1}(n-k)q_{k\alpha},\\
		\sigma_{i\alpha}-\sigma_{i+1,\alpha}=nq_{i\alpha}\in n\Z^+. \label{eqn:q_from_sigma}
	\end{align}
	
	The sixth is that \cref{eqn:sigma3,eqn:sigma1} are homogeneous, so not only is solving them over the ring of integers equivalent (via clearing denominators) to solving them over the field of rational numbers, but also we can (try to) do so by studying the projective variety that they define.\footnote{According to \cref{eqn:q_from_sigma}, it is not quite sufficient to solve \cref{eqn:sigma3,eqn:sigma1} over the integers; we must also take into account the restrictions
		\begin{equation}
			\sigma_{i\alpha}-\sigma_{i+1,\alpha}\in n\mathbb{Z}^+,i\in\{1,\dots,n-1\},\alpha\in\{1,\dots,m\}
		\end{equation}
		and will do so later on.}
	
	The seventh is that, because \cref{eqn:sigma1} are linear, this projective variety is a fairly simple one, namely a hypersurface (whose dimension is $m(n-1)-2$). Because of our passage to projective geometry, a point on this variety corresponds to an equivalence class containing infinitely many representations, whose $q_i$ or $\sigma_j$ (but not the Dynkin labels $m_i$) are all proportional to one another. 
	
	The eighth is that this cubic hypersurface features a number of projective linear subspaces (of dimension $(m(n-1)-2)/2$ for even $m(n-1)$ and $(m(n-1)-1)/2$ for odd $m(n-1)$). The presence of these linear subspaces is hardly serendipitous in itself, since they correspond precisely to the non-chiral reps (which are automatically anomaly-free). But their presence makes it easy to prove not only that the variety is rational (which, in layman's terms, means that we can solve this cubic equation), but also to construct, using the method of secants, an explicit parameterization of at least a dense subset (in the Zariski topology) of the solutions.
	
	The ninth is that our variety has a large degree of symmetry. Indeed, the equations \cref{eqn:sigma1,eqn:sigma3} that define it show immediately that its automorphisms contain a subgroup isomorphic to the semi-direct product $\underbrace{\left(S_n\times S_n\times\dots\times S_n\right)}_{m\text{ factors }} \rtimes S_m$, where the action of each $S_n$ factor is to permute the coordinate $\sigma_{i\alpha}$ for each $\alpha\in\{1,\dots,m\}$ and the action of the $S_m$ factor is to permute the tuples $\sigma_\alpha:=(\sigma_{1\alpha},\dots,\sigma_{n\alpha})$. This large symmetry has at least two benefits. One is that we can exploit it to produce an explicit parameterization of every solution (not just a dense subset of solutions as above). The other is that we can use it to turn integer solutions of \cref{eqn:sigma1,eqn:sigma3} that do not satisfy \cref{eqn:q_from_sigma} into ones that do, simply by permuting the $\sigma_{i\alpha}$, provided that for every $\alpha$, no two $\sigma_{i\alpha}$ are equal. This results in a very efficient computational algorithm for finding anomaly-free products of irreps.
	
	To begin describing all of this more explicitly, let us establish some notation. It will occasionally be convenient to work over the complex or real numbers as well as over the rationals, so we let $\Bbbk\in\{\C,\R,\Q\}$ denote such a field. 
	A point in the projective space $\Bbbk P^{nm-1}$ has homogeneous co-ordinates $[\sigma_{11}:\dots:\sigma_{n1}:\sigma_{12}:\dots:\sigma_{n2}:\dots:\sigma_{1m}:\dots:\sigma_{nm}]$ and is thus given by an equivalence class of $(\sigma_{11},\dots,\sigma_{nm})\in\Bbbk^{nm}$, where the equivalence relation is given by $(\sigma_{11},\dots,\sigma_{nm})\sim(\lambda\sigma_{11},\dots,\lambda\sigma_{nm})$ for any non-zero $\lambda\in\Bbbk$. The zero locus of \cref{eqn:sigma1,eqn:sigma3} defines a projective variety $V_{n,\otimes m}$ in $\Bbbk P^{nm-1}$.
	
	To deal with the restriction imposed by \cref{eqn:q_from_sigma}, we define the following four regions\footnote{These regions correspond to (semi-)algebraic sets, a notion which (for varieties over $\C$) has become important elsewhere in physics in the study of scattering amplitudes via so-called `positive geometry'.} on this variety:
	\begin{enumerate}
		\item The \emph{unorderable} region, for which $\exists\,i_1\neq i_2\in\{1,\dots,n\}$ such that $\sigma_{i_1\alpha}=\sigma_{i_2\alpha}$ for some $\alpha\in\{1,\dots,m\}$;
		\item Its complement, the \emph{orderable} region;
		\item For $\Bbbk\neq\C$, the \emph{weakly-ordered} region, for which either $\sigma_{1\alpha}\geq\sigma_{2\alpha}\geq\dots\geq\sigma_{n\alpha}$ or $\sigma_{1\alpha}\leq\sigma_{2\alpha}\leq\dots\leq\sigma_{n\alpha}$ for \emph{all} $\alpha\in\{1,\dots,m\}$;
		\item For $\Bbbk\neq\C$, the \emph{ordered} region, in which we replace the weak inequalities in the previous definition by strict ones.
	\end{enumerate}
	
	Before stating the results that are valid for general $n$ and $m$, we first consider the simplest non-trivial case, which is that of products of two irreps of $\mathfrak{su}_3$. Here our projective variety describes a surface in three dimensions, allowing us to draw pictures (at least over $\R$).
	
	\section{Binary products of $\mathfrak{su}_3$ \label{sec:bin3}}
	Since $\mathfrak{su}_2$ has no anomalous representations, the simplest non-trivial case is $\mathfrak{su}_3$, which of course describes the strong interactions in Nature. There, we currently only observe a non-chiral representation of $\mathfrak{su}_3$, made up of the six flavour of quarks and their antiquarks. But it is of interest in going beyond the Standard Model to ask whether we might in the future be able to observe fermions carrying chiral representations. For $m=1$, it is easy to see that there are no chiral irreps, because the equation describing the variety reduces to
	\begin{equation}
		\sigma_{11}\sigma_{21}(\sigma_{11}+\sigma_{21})=0,
	\end{equation}
	so we have exactly 3 rational points. Of these, only the point $\sigma_{21}=0$ is in the ordered region and its equivalence class consists of irreps of the form $(q_{11},q_{11})$, which are precisely the non-chiral irreps of $\mathfrak{su}_3$. We will now see however that we can find chiral anomaly-free representations for $m = 2$ (and, {\em ergo}, for all $m \geq 2$).
	
	For $m=2$, the variety $V_{3,\otimes2}$ can be described by the single equation
	\begin{equation}
		\sigma_{11}\sigma_{21}(\sigma_{11}+\sigma_{21})+\sigma_{12}\sigma_{22}(\sigma_{12}+\sigma_{22})=0.
	\end{equation}
	This describes a smooth cubic surface, so contains exactly 27 lines over $\Bbbk=\C$. Nine of these are given by
	\begin{equation}
		\sigma_{i1}+\sigma_{j1}=\sigma_{k1}=\sigma_{a2}+\sigma_{b2}=\sigma_{c2}=0,
	\end{equation}
	with $\{i,j,k\}=\{a,b,c\}=\{1,2,3\}$, and all of them are rational (that is, they also define lines in the variety over $\Q$). The remaining eighteen are given by
	\begin{equation}
		\sigma_{i1}+\omega\sigma_{a2}=\sigma_{j1}+\omega\sigma_{b2}=\sigma_{k1}+\omega\sigma_{c2}=\sigma_{i1}+\sigma_{j1}+\sigma_{k1}=0,
	\end{equation} 
	again with $\{i,j,k\}=\{a,b,c\}=\{1,2,3\}$, where $\omega$ is a cube root of $1$. Only the six lines with $\omega=1$ are real and rational. Thus, there are $9+6=15$ rational lines on $V_{3,\otimes2}$, 12 of which can be seen on the affine patch $\sigma_{31}+\sigma_{32}\neq0$ that we use for our Figures.\footnote{We chose this affine patch because it contains the ordered region.} On this affine patch, we define the coordinates
	\begin{align}
		x_1=-\frac{\sigma_{11}}{\sigma_{31}+\sigma_{32}},&&x_2=-\frac{\sigma_{21}}{\sigma_{31}+\sigma_{32}},&&x_3=-\frac{\sigma_{12}}{\sigma_{31}+\sigma_{32}}, \label{eqn:x123}
	\end{align}
	such that $\sigma_{22}=-(\sigma_{31}+\sigma_{32})(1-x_1-x_2-x_3)$ and that $V_{3,\otimes2}$ is described by
	\begin{equation}
		x_1x_2(x_1+x_2)+x_3(1-x_1-x_2-x_3)(1-x_1-x_2)=0.
	\end{equation}
	
\begin{figure}
		\centering
		\includegraphics[width=\linewidth]{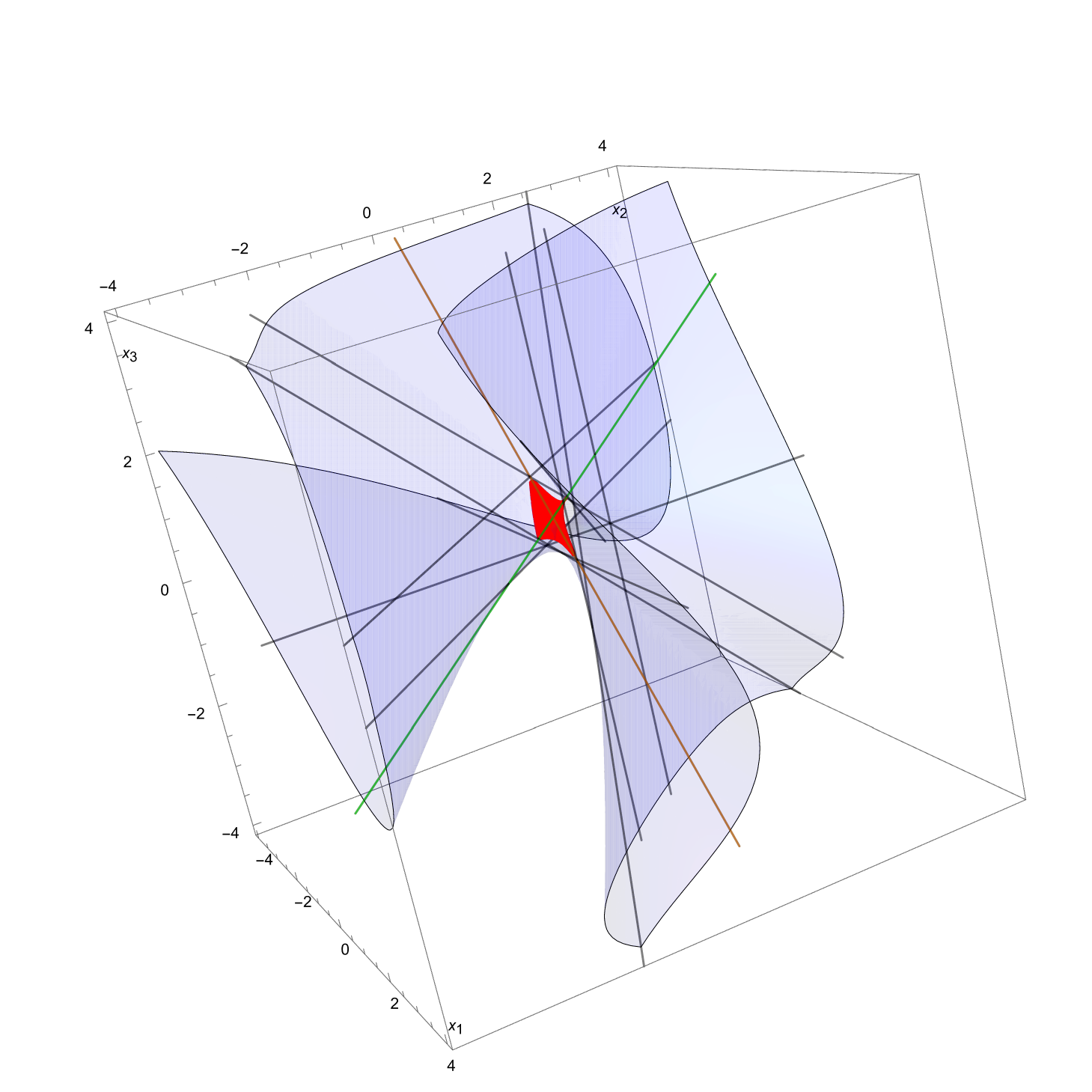}
		\caption{The variety $V_{3,\otimes2}$ in the affine patch $\sigma_{31}+\sigma_{32}\neq0$, along with its 12 rational lines. The ordered region is shaded in red. An ordered rational point on the orange line corresponds to a product of two non-chiral irreps, while one on the green line corresponds to the product of an irrep with its dual. The rational points contained in each of these real regions are dense in them, in the euclidean topology. The coordinates $x_1$, $x_2$ and $x_3$ are defined in~\cref{eqn:x123}.}
		\label{fig:v32productlines} 
	\end{figure}

	Fig.~\ref{fig:v32productlines} shows the variety, along with its rational lines.
	Two of these lines have special relevance for physics. Namely, the lines
	$\sigma_{11}+\sigma_{31}=\sigma_{21}=\sigma_{12}+\sigma_{32}=\sigma_{22}=0$ and $\sigma_{11}+\sigma_{32}=\sigma_{21}+\sigma_{22}=\sigma_{31}+\sigma_{12}=\sigma_{11}+\sigma_{21}+\sigma_{31}=0$, which are coloured orange and green respectively in \cref{fig:v32productlines}, intersect the ordered region, where they correspond to representations. The ordered rational points on the first line correspond to products of two non-chiral irreps, while those on the second line correspond to products of an irrep with its dual. Both classes of products are trivially anomaly-free and overall non-chiral; indeed, they comprise all non-chiral solutions to the anomaly cancellation for $n=3$, $m=2$. 
	Evidently, these points are dense in the underlying real region, which is diffeomorphic to a union of two copies of $\R^1$. We shall soon see, however, that the chiral anomaly-free representations are dense in the two-dimensional region given by the complement of this region in the ordered region, shown in red in the Figure. 
	In this sense, the non-chiral binary products are overwhelmed by the chiral ones.
	
	Just as in the case with the variety $V_{5,\otimes1}$ studied in \cite{Gripaios_2024}, there are disjoint (\emph{i.e.} skew) rational lines on $V_{3,\otimes2}$ (for example $\sigma_{11}+\sigma_{21}=\sigma_{31}=\sigma_{12}+\sigma_{22}=\sigma_{32}=0$ and $\sigma_{11}=\sigma_{21}+\sigma_{31}=\sigma_{12}=\sigma_{22}+\sigma_{32}=0$), and so the variety $V_{3,\otimes2}$ is rational, and we can use the method of secants described in \cite{Allanach_2020,Gripaios_2024} to generate at least a dense subset of the rational points on it.
	
	In brief, the method of secants can be described as follows. Given any two distinct points on a cubic, the line through those points must either intersect the cubic at a third point (which may coincide with one of the other two points), or must lie in the cubic. So by taking a pair of disjoint lines, we can construct a rational map, from $\Bbbk P^1 \times \Bbbk P^1 $ to the cubic, whose value is given by this third point. Since every point in the underlying projective space that is not on the disjoint lines ({\em ergo} every such point on the cubic) lies on exactly one line between the disjoint lines, it follows that one can construct an inverse to this map, exhibiting it as a birational map. (We refrain from giving an explicit description of the map here, since it is rather messy. The reader can easily construct it for themself, in analogy with the construction in \cite{Gripaios_2024}.) In fact, it is easy to see that the only points we miss are those on the disjoint lines themselves, or those on any third line that intersects both lines. These additional points are easily found, since we have already found all of the rational lines on $V_{3,\otimes2}$. 
	
	Just as for the case of $V_{5,\otimes1}$ studied in \cite{Gripaios_2024}, it also follows from this construction that the ordered rational points on $V_{3,\otimes2}$ are dense (not just in the Zariski topology but also in the euclidean topology) in the ordered real points; the analogous statement also holds for the orderable points.
	
\begin{figure}
		\centering
		\includegraphics[width=\linewidth]{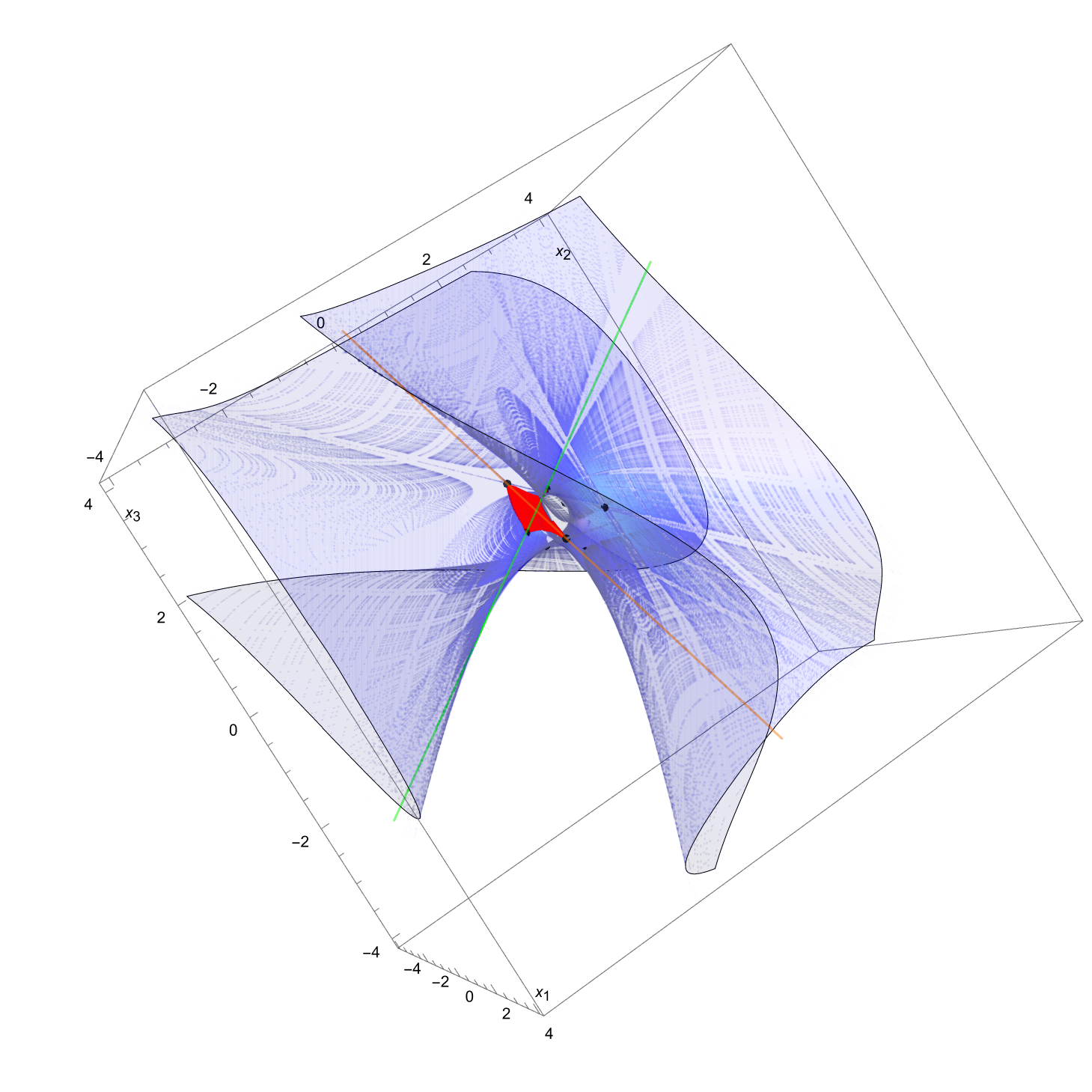}
		\caption{Rational points on the variety $V_{3,\otimes2}$ in the affine patch $\sigma_{31}+\sigma_{32}\neq0$ produced by a parameter scan. The red points are in the ordered region and the blue points are in the orderable region, so yield anomaly-free representations after applying a suitable automorphism.}
		\label{fig:v32productrationalpoints}
	\end{figure}

\begin{figure}
		\centering
		\includegraphics[width=\linewidth]{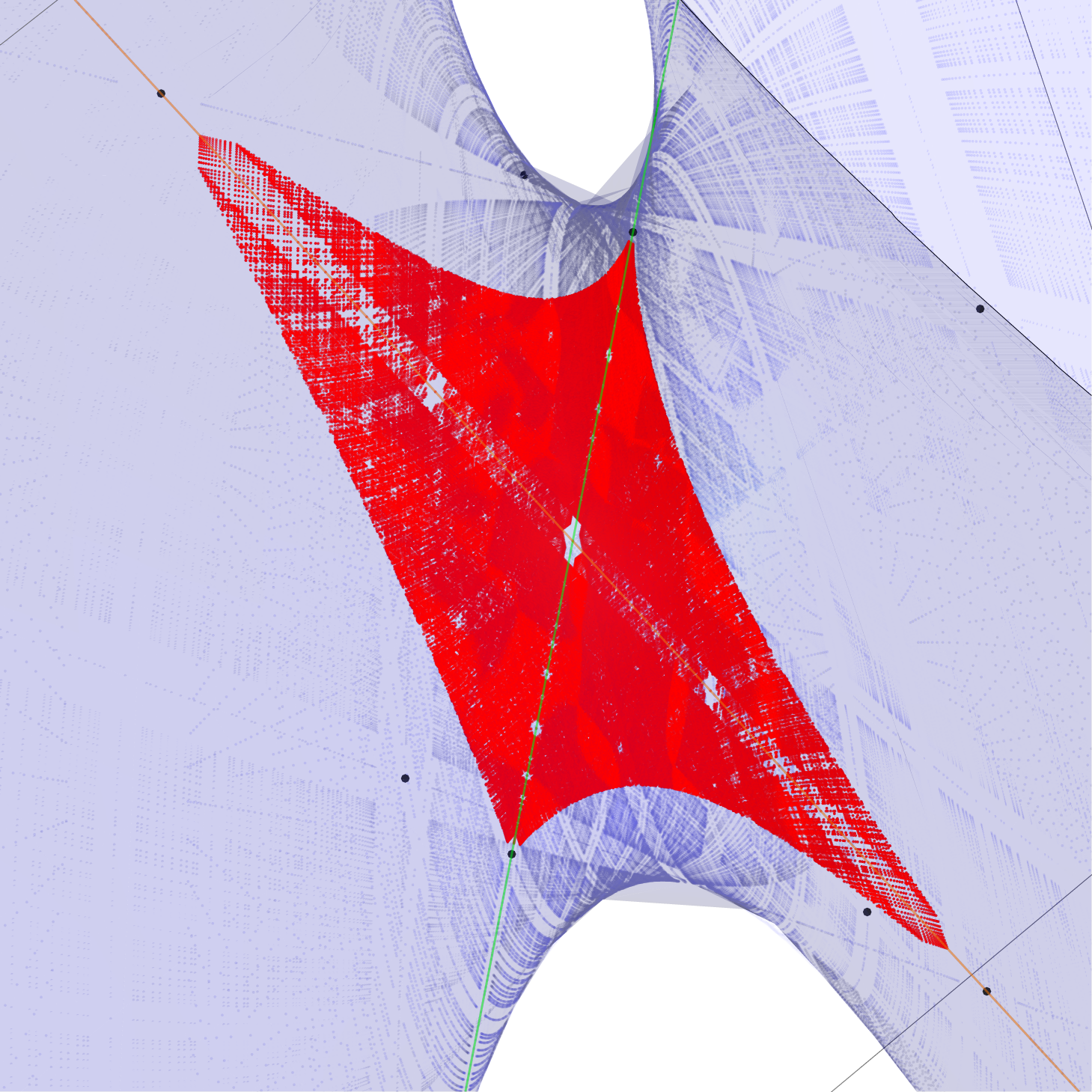}
		\caption{A close-up of the ordered region in Fig.~\ref{fig:v32productrationalpoints}.}
		\label{fig:v32productorderedregion}
	\end{figure}

	Fig.~\ref{fig:v32productrationalpoints} shows the rational points produced in this way by a scan over the parameters, while Fig.~\ref{fig:v32productorderedregion} shows a close-up of the ordered region.
	
\begin{figure}
		\centering
		\includegraphics[width=\linewidth]{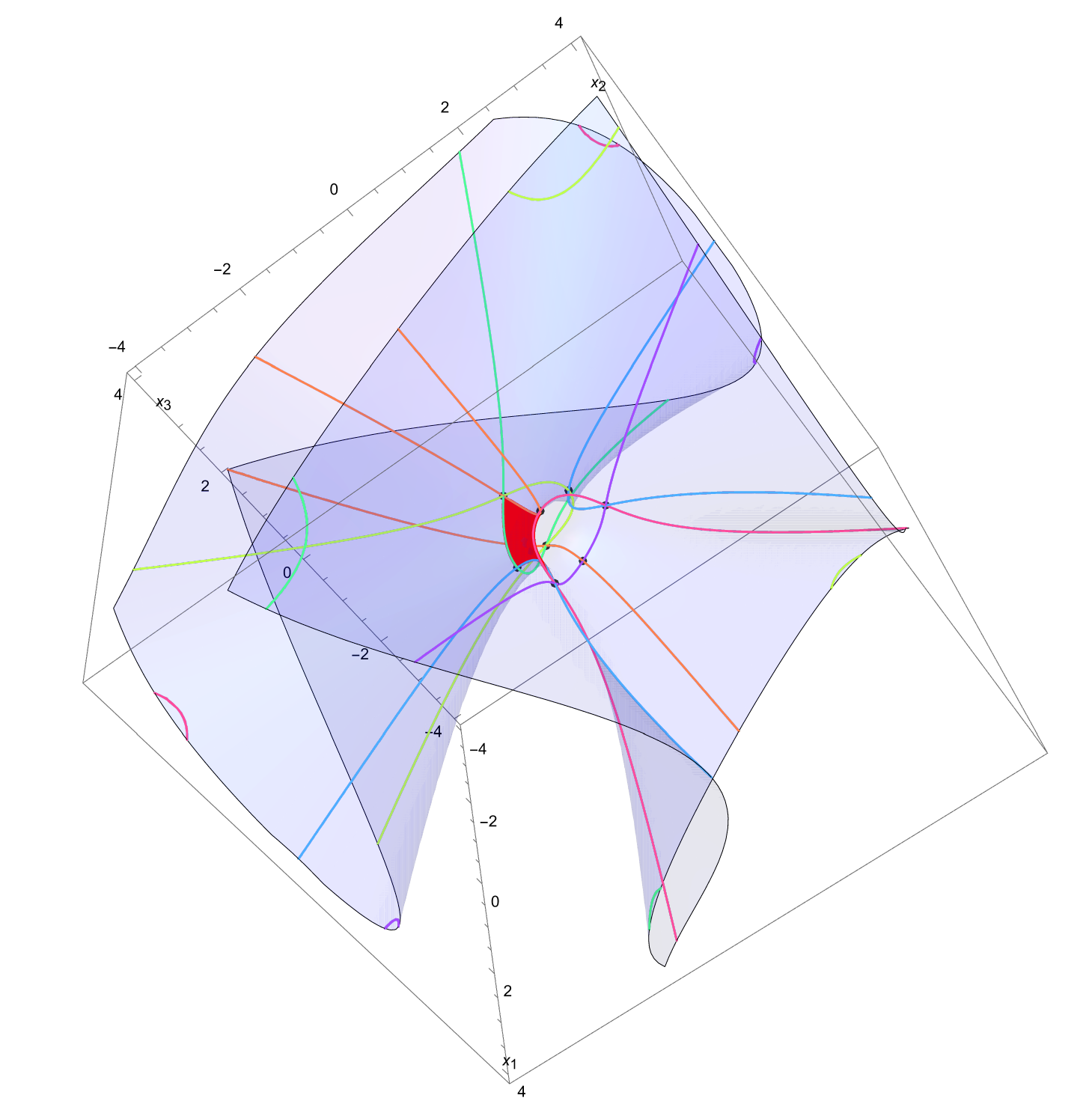}
		\caption{The six elliptic curves making up the unorderable region on $V_{3,\otimes2}$. Each curve has six rational points and there are 15 distinct rational points, of which eight are visible on this affine patch.} 
		\label{fig:v32productellipticcurves}
	\end{figure}

	Of the rational points produced by the method of secants in this way, we must reject the unorderable ones, since we cannot use them to construct representations. We will now show that (just as for $V_{5,\otimes1}$ in \cite{Gripaios_2024}), there are many fewer such points than expected. In particular, unlike the ordered and the orderable points, the unorderable rational points are not dense (in either topology) in the unorderable real points, but rather are finite in number. To see this, we observe that such points have $\sigma_{i_1j}=\sigma_{i_2j}$ for any $j\in\{1,2\}$ and $i_1\neq i_2\in\{1,2,3\}$. Thus, they are given by the $(S_3\times S_3)\rtimes\Z/2$ orbit of the curve
	\begin{equation}
		2(\sigma_{11})^3+\sigma_{12}\sigma_{22}(\sigma_{12}+\sigma_{22})=0,
	\end{equation}
	where $\sigma_{21}=\sigma_{11}$ and $\sigma_{31}=-2\sigma_{11}$. The part of this 
	curve with $\sigma_{11}=\sigma_{21}\geq\sigma_{31}$ and $\sigma_{12}\geq\sigma_{22}\geq\sigma_{32}$ belongs to the boundary of the weakly-ordered region on $V_{3,\otimes2}$. It is not hard to see that $[\sigma_{11}:\dots:\sigma_{32}]=[0:0:0:1:0:-1]$ is a rational point on the curve. Therefore, we have a smooth projective plane cubic equipped with a rational point, which defines an elliptic curve $C$. As with the case of $V_{5,\otimes1}$, we can calculate the Mordell-Weil group of the $\Bbbk$-rational points on $C$. On the one hand, over $\Bbbk=\R$, $C(\R)\cong U(1)$ because the cubic form has negative determinant. In other words, the $2\times {}^3C_2=6$ elliptic curves over $\R$ in the $S_3\times S_3\rtimes\Z/2$ orbit making up the real unorderable region each have one connected component containing infinitely many points. On the other hand, over $\Bbbk=\Q$ we can use the change of coordinates
	\begin{align}
		X=2\sigma_{11},&&Y=\sigma_{12}-\sigma_{22},&&Z=\sigma_{12}+\sigma_{22},
	\end{align}
	to show that $C$ is projectively equivalent to a Mordell elliptic curve of the form
	\begin{equation}
		y^2=x^3+1,
	\end{equation}
	whose Mordell-Weil group is isomorphic to $\Z/6$ \cite{Mordell_1969, Silverman_1992}; explicitly, the rational points on this curve are given by
	\begin{multline}
		C(\Q)=\{[0:0:0:1:-1:0],[1:1:-2:2:-1:-1],[0:0:0:1:0:-1],\\
		[-1:-1:2:1:1:-2],[0:0:0:0:1:-1],[1:1:-2:-1:2:-1]\},
	\end{multline}
	where the elements are ordered as the powers of a generator of $\Z/6$, beginning with the identity. Of each of these six points, three (the even powers of the generator) lie on three elliptic curves (for example, $[0:0:0:1:-1:0]$ satisfies $\sigma_{11}=\sigma_{21}=\sigma_{31}$, while the other three (the odd powers) lie on two elliptic curves (for example, $[1:1:-2:2:-1:-1]$ has $\sigma_{11}=\sigma_{21}$ and $\sigma_{22}=\sigma_{32}$). Thus, there are only $6\times 3/3+6\times 3/2=15$ unorderable $\Q$-rational points on $V_{3,\otimes2}$, eight of which are visible on the affine patch $\sigma_{31}+\sigma_{32}\neq0$. As promised, we see that the unorderable rational points are not dense in the unorderable real points, implying that our algorithm for finding representations is highly efficient. The curves and their rational points are shown in Fig.~\ref{fig:v32productellipticcurves}.
	
	A plot of the ordered real points on this affine patch of $V_{3,\otimes2}$ seems to suggest that, just as with the case for $V_5$, they form a connected region that is diffeomorphic to $\R^2$, though we do not prove it here.
	\section{More products of $\mathfrak{su}_n$ \label{sec:moreprod}}
	By way of generalizing the results for $n=3$, $m=2$ to arbitrary $n$ and $m$, we will prove the following:
	\begin{enumerate}
		\item The variety $V_{n,\otimes m}$ is rational over $\Q$, $\R$ and $\C$ (so we can parameterize the rational points);
		\item The (ordered, respectively orderable) rational points in $V_{n,\otimes m}$ are dense, in either the Zariski topology or the euclidean topology, in the (ordered, respectively orderable) real points;
		\item The ordered rational points that correspond to overall non-chiral products are dense (in the euclidean topology) in a union of subspaces, each of dimension $m(n-1)/2-1$ for odd $n$ or at most $mn/2-1$ for even $n$. Thus the chiral anomaly-free products overwhelm the non-chiral ones.
	\end{enumerate}
	We show that the variety $V_{n,\otimes m}$ (over $\Bbbk\in\{\Q,\R,\C\}$) is rational by generalizing the method of secants described for $n=3$, $m=2$ in the previous Section. The $m$ homogeneous linear constraints (\cref{eqn:sigma1}) and the cubic constraint (\cref{eqn:sigma3}) imply that the variety $V_{n,\otimes m}$ can also be viewed as a cubic hypersurface in $\Bbbk P^{m(n-1)-1}$ defined by
	\begin{equation}
		\sum_{j=1}^m\left[\sum_{i=1}^{n-1}\left(\sigma_{ij}\right)^3-\left(\sum_{i=1}^{n-1}\sigma_{ij}\right)^3\right]=0.
	\end{equation} 
	The analogue of the proof in \cite{Gripaios_2024} now implies that the method of secants generalizes as follows: instead of picking a pair of disjoint rational lines as in the $n=3$, $m=2$ case, we now need to find a pair of disjoint linear subspaces of $\Bbbk P^{m(n-1)-1}$ of dimensions $d_1=d_2:=(m(n-1)-2)/2$ if $m(n-1)$ is even and $d_1:=(m(n-1)-1)/2$, $d_2:=d_1-1$ if $m(n-1)$ is odd, such that the method of secants, including the step of permuting coordinates within each irrep factor to get ordered points from orderable ones, works as before. Note that $m(n-1)$ is even if $n$ is odd and vice versa, so these requirements do indeed reduce to those found in \cite{Gripaios_2024} for the case $m=1$. To get a pair of disjoint linear subspaces for every $n$ and $m$, we can assign to each of the $m$ irrep factors a pair of disjoint linear subspaces of the appropriate dimensions for the case $m=1$; examples of the latter can be found in \cite{Allanach_2020, Gripaios_2024}. To make this concrete, consider the example of $n=5$, $m=2$, where the homogeneous coordinates $[\sigma_{11}:\dots:\sigma_{51}:\sigma_{12}:\dots:\sigma_{52}]$ on two subspaces can be parameterized as
	\begin{align}
		\Gamma^1&=[k_1:k_2:0:-k_2:-k_1:l_1:l_2:0:-l_2:-l_1],\\
		\Gamma^2&=[0:m_1:-m_1:m_2:-m_2:0:n_1:-n_1:n_2:-n_2].
	\end{align}
	Thus, the proof in \cite{Gripaios_2024} generalizes for arbitrary $n$ and $m$: we always have a rational variety, where the birational map is given by the method of secants construction as described earlier. 
	
	Since the method of secants, which gives a manifestly continuous map with respect to either the Zariski topology or the usual euclidean topology on the real variety, works for our cubic hypersurfaces with arbitrary $n$ and $m$, it follows that on all of these hypersurfaces, the (ordered, respectively orderable) rational points are dense in the (ordered, respectively orderable) real points in either topology. At this point, we recall that for the case $n=3$, $m=2$, we observed from a plot of an affine patch of $V_{3,\otimes2}$ that the ordered region on it appears to be diffeomorphic to $\R^2$. Based on this, and based on what we obtained for $V_{n,\otimes1}$ in \cite{Gripaios_2024}, we conjecture that, in general, the ordered real region on $V_{n,\otimes m}$ is contractible and is of dimension $m(n-1)-2$.
	
	Assuming this conjecture holds, we can compare the density of chiral {\em vs.} non-chiral points. To do so, suppose that we have a partition $m=m_1+2m_2$, where $m_1$ and $m_2$ are both non-negative integers, such that we form a non-chiral $m$-fold product representation of $\mathfrak{su}_n$ from $m_1$ non-chiral irreps and $m_2$ pairs of dual chiral irreps. If $n$ is odd, such a representation can be seen to lie on a rational linear subspace of dimension $m_1(n-1)/2+m_2(n-1)-1=m(n-1)/2-1$, while if $n$ is even, it lies on a rational linear subspace of dimension $m_1n/2+m_2(n-1)-1=mn/2-m_2-1\leq mn/2-1$. Thus, we see that indeed the chiral products always overwhelm the non-chiral ones for $m=1$ and $n\geq5$, or $m\geq2$ and $n\geq3$.\footnote{In more detail, for odd $n$, we need $m(n-1)-2>m(n-1)/2-1$, which gives $m(n-1)>2$, so either $m=1$ and $n\geq5$, or $m\geq2$ and $n\geq3$, (both with $n$ odd); for even $n$, we need $m(n-1)-2>mn/2-1$, which gives $m(n-2)>2$, so either $m=1$ and $n\geq6$, or $m\geq2$ and $n\geq4$ (both with $n$ even). Putting these together, we obtain the claimed result.}
	
	We conclude this Section by listing in \cref{tab:tensor_product} some low-dimensional anomaly-free products for various $n$ and $m$ that we obtained by the method of secants. The representations are labelled by variables $q_{i\alpha}$ related to $\sigma_{i\alpha}$ by~\cref{eqn:q_from_sigma}. An implementation of the method of secants to obtain the results for $n=3$, $m=2$ as an example is included as an ancillary \texttt{Mathematica} notebook.
	
	There remains the question of whether the low-dimensional anomaly-free products in \cref{tab:tensor_product} are in fact the low{\em est}-dimensional ones. This is difficult to determine using the method of secants (which does not naturally order solutions in size), but can be verified by showing that product representations of lower dimension are anomalous. Since the dimensions increase monotonically with the Dynkin labels $m_i$ and since the weights are bounded below, this is a computation that can be carried out in a finite number of steps; however, we do not attempt to perform this calculation here.
	
	\begin{table}
		\caption{\label{tab:tensor_product} Some anomaly-free chiral products of $m$ irreducible representations of $\mathfrak{su}_n$ for some values of $n$ and $m$, with dimension below $10^5$.}
		\centering
		\begin{tabular}{LLLLLR}
			n&m&\bigotimes_{\alpha=1}^{m}(q_{1,\alpha},\dots,q_{n-1,\alpha})&& & \dim \\ \hline
			\rule{0pt}{3ex} 
			3&2 & (1,7)\otimes(7,4) & 28\times154&=& 4\,312 \\
			&& (2,7)\otimes(8,6) & 63\times336&=& 21\,168 \\
			&& (1,8)\otimes(12,11) & 36\times1518&=& 54\,468 \\ \hline
			\rule{0pt}{3ex} 
			3&3 & (2,1)\otimes(4,1)\otimes(4,5) & 3\times10\times90&=&2\,700 \\
			&& (1,4)\otimes(1,4)\otimes(5,2) & 10\times10\times35&=&3\,500 \\
			&&(1,2)\otimes(1,6)\otimes(6,3) & 3\times21\times81&=&5\,103 \\
			&&(1,2)\otimes(3,6)\otimes(6,2) & 3\times81\times48&=&11\,664 \\
			&&(1,3)\otimes(1,7)\otimes(8,7) & 6\times28\times336&=&56\,448 \\
			&&(1,3)\otimes(2,9)\otimes(9,3) & 6\times 99\times162&=&92\,228 \\ \hline
			\rule{0pt}{3ex} 
			4&2 & (1,4,3)\otimes(4,1,2) & 280\times70&=&19\,600 \\
			&& (1,1,5)\otimes(5,2,3) & 35\times875&=&30\,625 \\
			&& (4,1,1)\otimes(3,4,4) & 20\times2464&=&49\,280 \\
			&& (1,1,7)\otimes(7,1,3) & 85\times616&=&51\,744 \\
			&& (1,1,5)\otimes(4,5,2) & 35\times2310&=&80\,850
		\end{tabular}
	\end{table}

	\section{Phenomenology \label{sec:pheno}}
	
	As one can see, even going from $m=1$ to $m=2$ allows us to decrease the dimension of chiral anomaly-free representations by a factor of roughly 1000, bringing us much closer to the realm that seems reasonable, {\em a priori}, for phenomenology.
	
	Having representations that are small is desirable not just on the grounds of minimality, but also because it is a necessary condition to have a theory of physics at all. Indeed, a large matter representation
	in a gauge theory typically leads to a large (and positive) beta function for the gauge coupling, which in turn leads to the theory becoming strongly-coupled at energies not far above the mass of the matter, and thus a theory of physics that is valid only over a very short range of energy scales.
	
	Let us now discuss this in more detail. In general, the issue of whether a gauge theory is valid over a reasonable range of energies is both subjective and model-dependent, but there is one special class of theories in which is not the case, namely those where the beta function is negative, {\em i.e.} the theory is asymptotically-free. In such cases, it is clear that the theory is valid up to arbitrarily high energies. So let us begin by discussing this case.
	
	A first important observation is that the question of whether or not a representation $\rho$ is asymptotically-free is determined not by its dimension $D(\rho)$ but by its \emph{Dynkin index} $T_2(\rho)$, which for an irrep of $\mathfrak{su}_n$ is given by \cite{Okubo_1977}
	\begin{equation} \label{eqn:dynirrep}
		T_2(\rho)=\frac{D(\rho)}{2(n^2-1)}\left[\frac{1}{n^2}\sum_{i=1}^n\sigma_i^2-\frac{1}{12}n(n^2-1)\right],
	\end{equation}
	where the $\sigma_i$ are defined as in \cref{eqn:sigma_from_q}, and the index is normalised such that the defining representation with $m=(1,0,\dots,0)$ has $T_2=1/2$.
	
	Similar to the anomaly, we can relate the Dynkin index of a sum or product representation to those of its summands or factors according to
	\begin{align} 
		T_2(\rho_1\oplus\rho_2)&=T_2(\rho_1)+T_2(\rho_2),\label{eqn:dynsum}\\
		T_2(\rho_1\otimes\rho_2)&=D(\rho_2)T_2(\rho_1)+D(\rho_1)T_2(\rho_2),\label{eqn:dynprod}
	\end{align}
	and a theory of Weyl fermions transforming in a general representation $\rho$ of $\mathfrak{su}_n$ is asymptotically-free if
	\begin{equation}
		T_2(\rho)<\frac{11n}{2}.
	\end{equation}
	
	These formul\ae\ make it easy to determine, in principle, whether a given representation, expressed as a sum of irreps, is asymptotically-free. 
	But it is a rather more difficult problem to describe the set of all asymptotically-free representations. One reason for this difficulty is that even though the Dynkin index of an irrep in \cref{eqn:dynirrep} has the dimension as a factor, the former does not increase monotonically with the latter.\footnote{The Dynkin index does, however, increase monotonically with the Dynkin labels, as we will show elsewhere \cite{Gripaios_toappear}.} To give just one example
	the 21-dimensional $(q_1,q_2)=(6,1)$ irrep of $\mathfrak{su}_3$ has Dynkin index 35, which is higher than the Dynkin index of 25 of the 24-dimensional $(q_1,q_2)=(2,4)$ irrep. So one cannot simply ignore representations whose dimension is above some bound. 
	
	We remark that such an analysis was previously attempted in \cite{Eichten_1982}, but there were a number of errors. In particular, even the list of asymptotically-free irreps therein is incomplete. A hopefully more reliable analysis is in progress \cite{Gripaios_toappear}.
	
	For now, we content ourselves with showing that there {\em exists} a chiral representation that is a product of irreps and which is both asymptotically- and anomaly-free. Namely, the product representation of $\mathfrak{su}_7$ whose irrep factors are labelled by $(q_i) = (1,2,1,1,1,1)$ and $(q_i) = (1,1,1,1,1,2)$
	is anomaly-free according to \cref{eqn:sigma3}. It decomposes as the sum $(1,2,1,1,1,2)\oplus(2,1,1,1,1,1)$ (in terms of dimensions we have $21 \times 7 = 140 + 7$, while in terms of Young tableaux we have $\tiny\yng(2,2,1,1,1,1)\oplus\tiny\yng(1)=\tiny\yng(1,1)\otimes\overline{\tiny\yng(1)}$) and \cref{eqn:dynirrep,eqn:dynsum,eqn:dynprod} show that its Dynkin index is $28$; since $28<77/2$, it is also asymptotically-free. This representation, in its direct sum avatar, was previously also identified in \cite{Eichten_1982}.
	
	Once the analysis of \cite{Gripaios_toappear} has been carried out, it should be possible to give a more definitive statement about which product representations are both anomaly- and asymptotically-free.
	
	Finally, let us discuss the case of representations that lead to gauge theories that are not asymptotically-free, but neverthless have a reasonably small but positive beta function. Here things are much more subjective and model-dependent. By way of a benchmark, let us consider the Standard Model. Few would deny that it is a valid physical theory, but its hypercharge $\mathfrak{u}_1$ part is not asymptotically-free. Nevertheless, its Landau pole occurs at a fantastically high energy, so it could be valid over tens of orders of magnitude in energy (though let us hope it is not!). The size of the hypercharge beta function corresponds numerically to a Dynkin index of roughly $10^2$.
	
	Can we find anomaly-free product representations with Dynkin index in this ballpark? We can. For example, for $n=7$, the representation $\tiny\yng(2,2)\otimes\overline{\tiny\yng(2)}$ (of dimension $196\times28=5\,488$) has Dynkin index $2\,352$. Though an order of magnitude larger that the hypercharge value,
	this still gives plenty of room to build a perturbative gauge theory, especially when we consider that at larger values of $n$ it is arguably more appropriate to consider the relative size of the matter and gauge contributions to the beta function (here this factor is 61).
	
	For another example, for $n=14$, the anomaly-free representation $\tiny\yng(1,1,1,1)\otimes\overline{\tiny\yng(1,1)}$ (of dimension $1\,001\times91=91\,091$) has Dynkin index $16\,016$ (which is 208 times greater than the gauge contribution of $77$).
	
	Again, once the analysis of \cite{Gripaios_toappear} has been carried out, it should be possible to give a comprehensive analysis of which products of irreps are suitable for phenomenology.
	\section{The distribution of bounded rational points \label{sec:distro}}
	
	In this paper and its prequel \cite{Gripaios_2024}, we have seen that when anomaly-free chiral irreps (or products thereof) exist, they overwhelm the non-chiral ones, in the sense that they are dense (in the usual topology) in a real space of higher dimension. But we have also seen that such reps tend to have very large dimension. As such, they typically do not seem to be of much use for phenomenology, although the previous section shows that there are certainly exceptions.
	
	In fact, a related observation has already been made in a much more general context in diophantine geometry, as we now describe. Given a rational point $x$ on a projective variety $V$ in $\Q P^n$, we define its \emph{height} $H(x)$ by $\max(|x_0|,\dots,|x_n|)$, where the integers $x_0,\dots,x_n$ define a representative $(x_0,\dots,x_n)$ of $x$ and have no common prime factor. Given a subset $U \subset V$, we may then consider the number $N(U;B)$ of rational points of height at most $B\geq1$ in $U$.
	
	Suppose such a (projective) variety contains a (projective) line $L$. For large $B$, we have the asymptotic formula $N(L;B) \sim B^2$, a result known as Schanuel's theorem \cite{Schanuel_1979}.\footnote{A summary of the proof is as follows: It can be shown that calculating $N(\Q P^n, B)$ is equivalent to counting the number of \emph{integral} lattice points in an $(n+1)$-dimensional cube of side length proportional to $B$, and the number of such points scale as $B^{n+1}$. In particular, the increase in dimension from $n$ to $n+1$ is because a point in $n$-dimensional projective space corresponds to a line in $(n+1)$-dimensional affine space. In the limit of large $B$, and after considering all such cubes, the claimed asymptotic behaviour follows as the special case $n=1$.} On the other hand, a conjecture of Manin \cite{Franke_1989} asserts that (when $V$ is a smooth Fano variety whose rational points are Zariski dense) there exists a (non-empty) open set $U \subset V$ such that 
	\begin{equation}
		N(U;B) \sim B (\log B)^{\rho-1},
	\end{equation}
	where $\rho$ is the rank of the Picard group of $V$ (namely the set of isomorphism classes of line bundles on $V$ with group multiplication given by the tensor product of line bundles). Thus, there is a dense open set in $V$ (which clearly should not contain any lines, or more generally hyperplanes) that contains far fewer points than a line (or hyperplane), once we bound their size.
	
	Let us now consider how this can be applied to our set-up. 
	
	The variety $V_{n,\otimes m}$ is smooth if and only if $n$ is odd. Being a cubic hypersurface, it is Fano if and only if 
	\begin{equation}
		(n-1)m -1 \geq 3 \implies (n-1)m \geq 4.
	\end{equation} 
	We have seen that the rational points are always Zariski dense. Therefore, the conjecture applies for all odd $n$ and all $m$ except when $(n,m) = (3,1)$ (in which case we have seen that all points are nonchiral).\footnote{There are also versions of Manin's conjecture that apply to singular varieties (see {\em e.g.} \cite{Batyrev_1998}); we therefore expect that a similar conjecture applies for all $(n,m)$ except $(3,1)$ and $(4,1)$ (where there are also no nonchiral solutions).} In such cases, the height is given by removing any common factor from the shifted Dynkin labels $q_i$ and then taking the maximum. The points in the set $U$ correspond to chiral representations and their orbits under the automorphism group of the variety. So indeed, there is a general expectation that there are few chiral anomaly-free representations of bounded size, corresponding to the non-chiral representations.
	
	Unfortunately, Manin's conjecture has been proven in very few cases. Indeed, it has yet to be proven for any smooth cubic surface, including the varieties $V_{5,\otimes1}$ and $V_{3,\otimes2}$ considered here and in \cite{Gripaios_2024}.
	
	What we do have are corresponding upper and lower bounds. Specifically, we have an upper bound for any such variety $V$ featuring a line over $\mathbb{Q}$ given by \cite{Salberger_2015}
	\begin{equation}
		N(V\setminus \{\text{lines on $V$}\};B) \lesssim B^{12/7}
	\end{equation}
	and for a variety $V$ featuring three coplanar lines over $\Q$ given by \cite{HeathBrown_1997}
	\begin{equation}
		N(V\setminus \{\text{lines on $V$}\};B) \lesssim B^{4/3}. \label{eqn:4/3bound}
	\end{equation}
	
	We observe that we can find three coplanar lines on both $V_{5,\otimes1}$ and $V_{3,\otimes2}$, so that the stronger upper bound of \cref{eqn:4/3bound} holds. The former, which is projectively defined by
	\begin{equation}
		V_{5,\otimes1}:\sigma_1^3+\sigma_2^3+\sigma^3_3+\sigma_4^3-(\sigma_1+\sigma_2+\sigma_3+\sigma_4)^3=0,
	\end{equation}
	such that $\sigma_5:=-(\sigma_1+\sigma_2+\sigma_3+\sigma_4)$, contains the lines
	\begin{align}
		L_1:\sigma_1+\sigma_2=\sigma_3+\sigma_4=\sigma_5=0,\\
		L_2:\sigma_1+\sigma_2=\sigma_3+\sigma_5=\sigma_4=0,\\
		L_3:\sigma_1+\sigma_2=\sigma_3=\sigma_4+\sigma_5=0,
	\end{align}
	which all lie on the plane $\sigma_1+\sigma_2=0$. The latter is the set of projective solutions to the equation
	\begin{equation}
		V_{3,\otimes2}:\sigma_{11}\sigma_{21}(\sigma_{11}+\sigma_{21})+\sigma_{12}\sigma_{22}(\sigma_{12}+\sigma_{22})=0,
	\end{equation}
	where we have made the substitutions $\sigma_{31}=-(\sigma_{11}+\sigma_{21})$ and $\sigma_{32}=-(\sigma_{12}+\sigma_{22})$. In a similar way, it is evident that the plane $\sigma_{11}+\sigma_{21}=\sigma_{31}=0$ contains the three lines
	\begin{align}
		L'_1:\sigma_{11}+\sigma_{21}=\sigma_{31}=\sigma_{12}+\sigma_{22}=\sigma_{32}=0,\\
		L'_2:\sigma_{11}+\sigma_{21}=\sigma_{31}=\sigma_{12}+\sigma_{32}=\sigma_{22}=0,\\
		L'_3:\sigma_{11}+\sigma_{21}=\sigma_{31}=\sigma_{12}=\sigma_{22}+\sigma_{32}=0.
	\end{align}
	Thus, we conclude that chiral representations of bounded size are, asymptotically, rarer than non-chiral representations, at least in these two examples.
	
	Manin's relation has also been shown by Slater and Swinnerton-Dyer to be a lower bound if the smooth cubic surface contains two disjoint rational lines \cite{Slater_1998}; this is the case for both $V_{5,\otimes1}$ and $V_{3,\otimes 2}$, so (at least asymptotically) there are nevertheless plenty of chiral anomaly-free representations available for us to do physics with. To see just how many, let us close with a computation of $\rho$ in both cases. To do so, we follow \cite{Frei_2016}: given a smooth cubic surface with two disjoint lines in 3-dimensional projective space over any field of characteristic zero with co-ordinates $[x_0:x_1:x_2:x_3]$, we can do linear transformations to put the lines at $x_0 = x_1 = 0$ and $x_2 = x_3 = 0$. It follows from the second of these that the cubic can be written in the form
	\begin{gather}
		a(x_0,x_1)  x_2^2 + d(x_0,x_1) x_2 x_3 + f(x_0,x_1) x_3^2 + b(x_0,x_1) x_2 + e(x_0,x_1) x_3,
	\end{gather}
	where $a,d,f$ are linear homogeneous and $b,e$ are quadratic homogeneous functions. The rank $\rho$ is then given by $r+2$ where $r$ is the number of irreducible factors over the ground field of the quintic discriminant
	\begin{gather}
		\Delta =  ae^2 - bde + fb^2.
	\end{gather}
	For $V_{5,\otimes 1}$, we do the change of variables
	\begin{align}
		\sigma_1=-x_1-x_3,&&\sigma_2=x_0+x_1+x_3,&&\sigma_3=-x_2-x_3,&&\sigma_4=x_1+x_2+x_3,
	\end{align}
	such that the lines $L_1$ and $L_3$ have $x_0=x_1=0$ and $x_2=x_3=0$ respectively, to obtain
	\begin{gather}
		\Delta =  27x_0x_1(x_0+x_1)(x_0^2 + 3x_0x_1 + x_1^2)
	\end{gather}
	which has five reducible factors over $\C$ and $\R$, but only 4 over $\Q$, to conclude that $\rho = 6$.
	
	For $V_{3,\otimes 2}$, we do the change of variables
	\begin{align}
		\sigma_{11}=x_2,&&\sigma_{21}=x_0-x_1,&&\sigma_{12}=x_3,&&\sigma_{22}=x_1-x_3,
	\end{align}
	which maps $L'_1$ to the line $x_0=x_1=0$ and the line
	\begin{equation}
		L'_4:\sigma_{11}=\sigma_{21}+\sigma_{31}=\sigma_{12}=\sigma_{22}+\sigma_{32}=0
	\end{equation}
	to $x_2=x_3=0$, to obtain
	\begin{gather}
		\Delta =  -x_0x_1(x_0+x_1)(x_0^2 -x_0x_1 + x_1^2)
	\end{gather}
	which has five reducible factors over $\C$, but only 4 over $\R$ and $\Q$, to conclude again that $\rho = 6$. 
	
	\acknowledgments{This work has been partially supported by STFC consolidated grants ST/T000694/1 and ST/X000664/1, a JSPS Invitational Fellowship, and a Trinity-Henry Barlow Scholarship.}
	\bibliography{more_varieties_references} 
\end{document}